\documentclass[fleqn,11pt,twoside]{article}

\usepackage{amsthm,amssymb, color, xcolor, epsfig, graphics, subfigure}
\usepackage{amsmath, graphicx, latexsym, lscape}

\makeatletter
\newcommand{\copyrightnote}[2]{{\renewcommand{\thefootnote}{}
 \footnotetext{\small\it
\begin{flushleft}
 \copyright \ #1   #2  
\end{flushleft}}}}

\newcommand{\Name}[1]{\begin{flushleft}
                       \LARGE \bf #1
                       \end{flushleft}\vspace{-3mm}}

\newcommand{\Author}[1]{\begin{flushleft}
                       \it #1 \end{flushleft}}

\newcommand{\Address}[1]{\begin{flushleft}
                        \it #1 \end{flushleft}}

\newcommand{\Date}[1]{\begin{flushleft}
                      \small  \it #1 \end{flushleft}}

\newcommand{\evenhead}{Author \ name}
\newcommand{\oddhead}{Article \ name}

\renewcommand{\@evenhead}{
\hspace*{-3pt}\raisebox{-15pt}[\headheight][0pt]{\vbox{\hbox to \textwidth
{\thepage \hfil \evenhead}\vskip4pt \hrule}}}
\renewcommand{\@oddhead}{
\hspace*{-3pt}\raisebox{-15pt}[\headheight][0pt]{\vbox{\hbox to \textwidth
{\oddhead \hfil \thepage}\vskip4pt\hrule}}}
\renewcommand{\@evenfoot}{}
\renewcommand{\@oddfoot}{}

\long\def\@makecaption#1#2{%
  \vskip\abovecaptionskip
  \sbox\@tempboxa{\small \textbf{#1.}\ \ #2}%
  \ifdim \wd\@tempboxa >\hsize
    {\small \textbf{#1.}\ \ #2}\par
  \else
    \global \@minipagefalse
    \hb@xt@\hsize{\hfil\box\@tempboxa\hfil}%
  \fi
  \vskip\belowcaptionskip}

\newcommand{\JNMPnumberwithin}[3][\arabic]{%
  \@ifundefined{c@#2}{\@nocounterr{#2}}{%
    \@ifundefined{c@#3}{\@nocnterr{#3}}{%
      \@addtoreset{#2}{#3}%
      \@xp\xdef\csname the#2\endcsname{%
        \@xp\@nx\csname the#3\endcsname .\@nx#1{#2}}}}%
}

\newcommand{\resetfootnoterule} {
  \renewcommand\footnoterule{%
  \kern-3\p@
  \hrule\@width.4\columnwidth
  \kern2.6\p@}
}

\renewcommand{\footnoterule}{}

\makeatother

\theoremstyle{definition}

\begin{document}

\renewcommand{\evenhead}{ {\LARGE\textcolor{blue!10!black!40!green}{{\sf \ \ \ ]ocnmp[}}}\strut\hfill 
C Rogers and S Carillo
}
\renewcommand{\oddhead}{ {\LARGE\textcolor{blue!10!black!40!green}{{\sf ]ocnmp[}}}\ \ \ \ \  
A Cuspon equation and Painlev\'e symmetry reduction
%: Exact Solution via Painleve' Symmetry Reduction
}

%%%% Matter for the first page 
\thispagestyle{empty}
\newcommand{\FistPageHead}[3]{
\begin{flushleft}
\raisebox{8mm}[0pt][0pt]
{\footnotesize \sf
\parbox{150mm}{{\textcolor{blue!10!black!40!green}{{\bf Open Communications in Nonlinear Mathematical Physics}}}
\ \ {Special Issue: Hietarinta}, 2026\\[0.1cm]
\strut\hfill 
ocnmp:18295
pp #2\hfill {\sc #3}}}\vspace{-13mm}
\end{flushleft}}

\FistPageHead{1}{\pageref{firstpage}--\pageref{lastpage}}{ \ \ }

\strut\hfill

\strut\hfill

\copyrightnote{Colin Rogers and Sandra Carillo
Distributed under a Creative Commons Attribution 4.0 International License}

\begin{center}
{\bf {\large A Special OCNMP Issue in Honour of Jarmo Hietarinta}}\\[0.2cm]
{\bf {\large on the Occasion of his 80th Birthday}}
\end{center}

\smallskip

\Name{Moving Boundary Problems for a Cuspon Equation and Reciprocal Associates: Exact Solution via Painlev\'e Symmetry Reduction}

\Author{Colin Rogers}

\Address{University of New South Wales, Australia}

\Author{Sandra Carillo}

\Address{Dipartimento Scienze di Base e Applicate per l'Ingegneria,{``Sapienza''} Universit\`a di Roma,  Rome, Italy \& Gr. Roma1, IV - Mathematical Methods in NonLinear Physics,  National Institute for Nuclear Physics (I.N.F.N.), Rome, Italy}

\Date{Received May 28, 2026; Accepted June 7, 2026}

\setcounter{equation}{0}

\smallskip

\noindent
{\bf Citation format for this Article:}\newline
Colin Rogers and Sandra Carillo,
Moving boundary problems for a Cuspon equation and reciprocal associates: exact solution via Painlev\'e symmetry reduction,
{\it Open Commun. Nonlinear Math. Phys.}, Special Issue:\,Hietarinta, ocnmp:18295, \pageref{firstpage}--\pageref{lastpage}, 2026.

\strut\hfill

\noindent
{\bf The permanent Digital Object Identifier (DOI) for this Article:}\newline
{\it 10.46298/ocnmp.18295}
\strut\hfill

\begin{abstract}
\noindent 
Here classes of moving boundary problems of Stefan-type for both an established nonlinear evolution
equation of cuspon theory and novel reciprocally  linked solitonic equations are shown to be solvable via Painlev\'e II
symmetry reduction.
\end{abstract}

\label{firstpage}

%%%% The Article text starts here

\section{Introduction}

Moving boundary problems of Stefan-type have had extensive application in continuum mechanics, notably in connection  with change of phase in a non-linear heat conduction context and liquid transport through porous media in soil mechanics \cite{ref1,ref2,ref3}  and \cite{ref4} (see also references cited therein).
In modern soliton theory, classes of moving boundary problems for the  canonical Dym equation \cite{ref5} and reciprocal associates are derived in \cite{ref6} which admit exact solutions via Painlev\'e II symmetry reduction.
This investigation was originally motivated,  in part, by the classical
Saffman-Taylor problem with surface tension \cite{ref7}.
The occurrence of the  solitonic Dym equation  in the Hele-Shaw theory was elucidated in \cite{ref8}.
Moving boundary problems  for a range of solitonic equations have been shown to admit exact solution via Painlev\'e II symmetry reduction
\cite{ref9,ref10,ref11,ref12,ref13,ref14}.
{{Moving boundary problems of Stefan-type amenable to exact solution have been recently studied in 
\cite{RogersBriozzo2025a, RogersBriozzo2025b, RogersAmster2026}.}}
Reciprocal-type transformations had their origin in the derivation of invariance properties of conservation laws in homentropic gas dynamics in  \cite{ref15} and were subsequently shown in  \cite{ref16} to constitute particular B\"acklund transformations   \cite{ref17,ref18}. 
In modern soliton theory, reciprocal transformations were introduced  in  \cite{ref19} 
and associated with admitted conservation laws.
Therein, these were  conjugated with the action of the classical Bianchi permutability theorem associated with invariance of the soliton system under a B\"acklund transformation.
{{The important role of B\"acklund Transformations in the construction of new solutions and in revealing new invariance both in the study of soliton equations, both in the commutative and  in the non-commutative cases are given in \cite{SCCS-OCNMP2024}(and references therein).}}  Thereby, multi-soliton solutions can be generated iteratively in an algorithmic manner. In  \cite{ref20}, reciprocal transformations were applied to link the canonical AKNS and WKI inverse scattering schemes of \cite{ref21} and \cite{ref22}, respectively. The linkage of certain classes of 1+1-dimensional solitonic hierarchies via reciprocal transformations have been detailed  in
\cite{ref23,ref24,ref25}. Reciprocal transformations in 2+1-dimensions, as originally introduced  in 
\cite{ref26},  have been applied to connect the  Kadomtsev-Petviashvili,  2+1-dimensional Dym and modified Kadomtsev-Petviashvili solitonic  hierarchies in  \cite{ref27}. 

In \cite{ref28}, a novel  nonlinear evolution equation descriptive of certain cuspon and periodic cuspon phenomena was obtained and notably, in particular, a Lax pair was derived.
Here, classes of moving boundary problems of Stefan-type, both for this cuspon equation and integrable extensions linked by reciprocal transformations, are shown to be solvable via Painlev\'e II symmetry reduction.

\section{A Class of Reciprocal Moving Boundary Problems for a \\ Solitonic Cuspon Equation: Painlev\'e II Symmetry Reduction} 
The reciprocal transformation
\begin{equation}
dx^* = v \, dx + ( - v_{xx} + 3v^2) \, dt, \quad t^* = t
\tag{2.1}
\end{equation}
with  compatibility condition for the canonical solitonic Korteweg-de Vries equation 
\begin{equation}
v_t - 6vv_x + v_{xxx} = 0
\tag{2.2}
\end{equation}
with $v = -1/m^*$ yields
\begin{equation}
dx = m^* \, dx^* + \left( \frac{1}{2} \frac{\partial^2}{\partial{x^*}^2} \left( \frac{1}{m^{*2}} \right) - \frac{3}{m^* }\right) dt^*.
\tag{2.3}
\end{equation}
The latter has compatibility condition 
\begin{equation}
m_t^* = \frac{1}{2} \left( \frac{1}{m^{*}} \right)_{x^*x^*x^*} - 3 \left( \frac{1}{m^*} \right)_{x^*},
\tag{2.4}
\end{equation}
namely, the solitonic cuspon  equation in \cite{ref28}.
In  \cite{ref13}, 
exact solutions of a class of  Korteweg-de Vries moving boundary problems has been solved via application of  the Miura transformation
\begin{equation}
v = u_x + u^2
\tag{2.5}
\end{equation}
which connects the KdV equation (2.2)  to the mKdV equation
\begin{equation}
u_t - 6u^2u_x + u_{xxx} = 0.
\tag{2.6}
\end{equation}
This important link may be derived in the context of 
a class of  classical  B\"acklund transformations due to to Clairin \cite{ref17}. 
 In  \cite{ref13}, application  was made of  Painlev\'e II symmetry reduction to solve a class of
 KdV moving boundary problems of Stefan-type problems governed by the system
 \begin{equation*}
v_t - 6vv_x + v_{xxx} = 0 ~~,~~ 0<x< S(t)= \gamma (t+a)^{1/3}~, t>0
\end{equation*}
\begin{equation}
\left. \begin{array}{l}
v_{xx} - 3v^2 = L_m S^i \dot{S} \\
v = P_m S^j
\end{array} \right\} \text{ on } x = S(t), \quad t > 0
\tag{2.7}
\end{equation}
$$(v_{xx} - 3v^2)|_{x=0} = H_0(t+a)^k, t > 0, $$ $$S(0) = S_0~.$$ The mKdV equation (2.6) may be be shown to admit  a 
 Painlev\'e II symmetry reduction with 
\begin{equation}
u = (t+a)^p \Psi(\xi), \quad \xi ={ x\over {(t+a)^q}}.
\tag{2.8}
\end{equation}
Thus, on substitution of the latter representation into  (2.6) there results 
\begin{equation}
p \Psi - q \xi \Psi' - 6(t+a)^{2p-q+1} \Psi^2 \Psi' + (t+a)^{-3q+1} \Psi''' = 0
\tag{2.9}
\end{equation}
whence $p = -1/3, q = 1/3$ and $\Psi(\xi)$ is governed by
\begin{equation}
\Psi''' - 6\Psi^2 \Psi' - \frac{1}{3} (\xi \Psi)' = 0 
\tag{2.10}
\end{equation}
so that, on integration, 
\begin{equation}
\Psi'' - 2\Psi^3 - \frac{1}{3} \xi \Psi = \kappa^*~,~ \kappa^*\in{\mathbf{R}}.
\tag{2.11}
\end{equation}
On introduction of the scalings $\Psi = \delta w$, $\xi = \epsilon z$, into the latter there results the classical Painlev\'e II equation \begin{equation}
w_{zz} = zw^3 + zw + \alpha~, 
\tag{2.12}\end{equation}
 with parameter $\alpha=  \kappa^* \epsilon^2/ \delta$.
Under the Miura transformation (2.5), the class of solutions 
\begin{equation}
v = (t+a)^{-2/3} (\Psi'(\xi) + \Psi^2(\xi)) := (t+a)^{-2/3} \Lambda(\xi)
\tag{2.13}
\end{equation}
 of  the KdV equation  (2.2) is obtained wherein  $ \Psi(\xi)$ is given by the scaled version (2.11) of the  Painlev\'e II equation.
The {{following}} KdV  moving boundary conditions {{are considered. In detail}} :
\begin{description}
\item[I ] {\begin{equation*}
v_{xx} - 3v^2 = L_m S^i \dot{S} \text{ on } x = S(t)=\gamma (t+a)^{1/3}~, \quad t > 0~.
\end{equation*}
Insertion of the relation (2.13) yields
\begin{equation}
(t+a)^{-4/3} (\Lambda''(\gamma) - 3\Lambda^2(\gamma)) = \frac{1}{3} L_m \gamma^{4i/3} (t+a)^{(i-2)/3}.
\tag{2.14}
\end{equation}
where $i=-2$ on alignment of the powers in $t+\alpha$
%The conclusion $i=-2$ comes from comparison of the powers of $\eta^\alpha$ %.
$i = -2$
and % together with 
\begin{equation}
L_m = 3\gamma^{-4i/3}  [\Lambda''(\gamma) - 3\Lambda^2(\gamma)].
\tag{2.15}
\end{equation}}

\item[II ]{
\begin{equation*}
v = P_m S^i \dot{S} ~~~\text{ on }~~ x = S(t)=\gamma (t+a)^{1/3}~, \quad t > 0~.
\end{equation*}
 This requires $j = -2$ 
and 
\begin{equation}
P_m = \gamma^2 \Lambda(\gamma) = \gamma^2 [\Psi'(\gamma) + \Psi^2(\gamma)].
\tag{2.16}\end{equation}
}
\item[III ]{
\begin{equation*}
(v_{xx} - 3v^2)\vert_{x=0} =H_0 (t+a)^k  
~, \quad t > 0~.
\end{equation*}
This boundary condition yelds $k=-4/3$ together with
\begin{equation}
H_0=\Lambda''(0) -3 \Lambda^2(0)~.
\tag{2.17}\end{equation}
}
\end{description}

The moving boundary problems for the cuspon equation (2.4) under the reciprocal transformation (2.1) 
with $ v= {1}/{m^*}$ become: 

\begin{equation*}
m_t^* = \frac{1}{2} \left( \frac{1}{m^{*2}} \right)_{x^*x^*x^*}
- 3 \left( \frac{1}{m^*} \right)_{x^*},
\qquad
x^*\vert_{x=0 }< x^* < x^*\vert_{x=S(t)}
:= S^*(t^*), \quad t^*>0.
\end{equation*}

\begin{equation}
\left.
\begin{array}{l}
\displaystyle{
\frac{1}{2} \left( \frac{1}{m^{*2}} \right)_{x^*x^*}
- 3 \left( \frac{1}{m^*} \right)
=
m^* L_m S^i \dot{S}}
\\[2ex]

\displaystyle{
 \frac{1}{m^*} 
=
P_m S^j}
\end{array}
\right\}
\qquad
\text{on } x^* = S^*(t^*), \quad t^*>0
\tag{2.18}
\end{equation}

\begin{equation*}
\left[
\frac{1}{2} \left( \frac{1}{m^{*2}} \right)_{x^*x^*}
- 3 \left( \frac{1}{m^*} \right)
\right]
\Bigg\vert_{{x^*}\vert_{x=0}}
=
m^*\big|_{x^*\vert_{x=0}}
H_0 (t+a)^k,
\qquad t^*>0
\end{equation*}

\begin{equation*}
S^*(0)=S_0^*.
\end{equation*}
In the preceding, $x^* = S^*(t^*)$ is the reciprocal moving boundary obtained by application of (2.1) so 
$x = S(t) = \gamma(t+a)^{1/3}.$
 Thus,
\begin{equation}
\begin{array}{l}
\displaystyle{
dx^* |_{x=S(t)} = \left[v dx - (v_{xx} - 3v^2) dt\right] |_{x=S(t)} }\\ \\
\displaystyle{\,\,\,\qquad\quad\quad= (P_m S^j \dot{S} + L_m S^i \dot{S}) dt}
\end{array}\tag{2.19}
\end{equation}
 wherein $i=j=-2$.
Accordingly,
\begin{equation}
dx^* |_{x=S(t)} = \left( P_m +L_m\right) {(1/ 3 \gamma) } (t^*+a)^{-4/3}  dt
\tag{2.20}\end{equation}
so that
\begin{equation}
S^*(t^*) = \gamma^* (t^*+a)^{-1/3} + \delta^*~~,~~\gamma^* , \delta^* \in{\mathbf{R}}
\tag{2.21}
\end{equation}
with $\gamma^* = -\gamma^{-1}(P_m + L_m)$.
The associated reciprocal initial boundary condition becomes:
\begin{equation}
S^*(t^*) = \gamma^* a^{-1/3} + \delta^*~~.
\tag{2.22}
\end{equation}
In addition,
\begin{equation}
dx^* |_{x=0} =  (-v_{xx} + 3v^2) dt  |_{x=0}= H_0 (t^*+a)^k  dt
\tag{2.23}\end{equation}
with $k=-1/3$, whence
\begin{equation}
x^* |_{x=0} = -3 H_0 (t^*+a)^{-1/3} + \epsilon^*~~,~~ \epsilon^* \in{\mathbf{R}}.
\tag{2.24}
\end{equation}
Thus, the region $$ x^*\vert_{x=0 }< x^* < x^*\vert_{x=S(t)}:= S^*(t^*)~,$$    reciprocally associated with $0 < x < S(t)$, is given by  
\begin{equation}
-3 H_0 (t^*+a)^{-1/3} + \epsilon^* < x^* < \gamma^* (t^*+a)^{-1/3} + \delta^*~.
\tag{2.25}
\end{equation}

It is remarked that moving boundary problems constrained by a pair of time-dependent boundaries occur, in particular, both in the context of resonant   nonlinear Schr\"odinger boundary analysis   \cite{ref29} and certain Stefan-type problems in the context of nonlinear heat conduction incorporating a source 
 term  \cite{ref30}.

\section{An Extended Solitonic Cuspon Equation: \\ Reciprocal Moving Boundary Problems} 
In \cite{ref31}, a novel solitonic extension of the cuspon equation (2.4) was introduced, namely:
\begin{equation}
m_t^* = \frac{1}{2} (m^{*-2})_{x^*x^*x^*} + \delta^* (m^{*-1})_{x^*} + \epsilon^* (m^{*-2})_{x^*}~~,~~ \delta^*, \epsilon^* \in{\mathbf{R}}
\tag{3.1}
\end{equation}
which is linked via a reciprocal transformation to the canonical solitonic Gardner equation 
\begin{equation}
v_\tau + 6v(1 -   v)v_y + v_{yyy} = 0.
\tag{3.2}
\end{equation}
The latter has diverse physical applications, notably in plasma physics  \cite{Ruderman2008}, 
optical lattice theory \cite{Wadati1975},
nonlinear wave, propagation phenomena in both hydrodynamics \cite{Grimshaw2010} and elastodynamics \cite{Coclite2021}.
On application to (3.2) of the reciprocal transformation 
\begin{equation}
dx^* = v \, dy - [v_{yy} + 3 v^2 - 2 v^3] d\tau, \quad t^* = \tau
\tag{3.3}
\end{equation}
with $v= { {1}/{m^{*}}}$, there results
\begin{equation}
dy=m^*dx^*+\left[\frac12\left(m^{*-2}\right)_{x^*x^*}+3m^{*-1}-2m^{*-2}\right]dt^*,
\tag{3.4}
\end{equation}
with compatibility condition the extended cuspon equation (3.1) with
parameters \hbox{\(\delta^*= 3,~ \beta = -2\).}

\section*{A Class of Moving Boundary Problems} 
 It was recently established in \cite{ref31}
 that a class of nonlinear moving
boundary problems for the Gardner equation (3.2) admits exact solution via
a Painlev\'e II symmetry reduction on application of a mKdV connection.
This class was determined by the nonlinear system

\begin{equation*}
v_\tau+6v(1-v)v_y+v_{yyy}=0,
\qquad
\frac{3\tau}{2}<y<\gamma(\tau+a)^{1/3}+\frac{3\tau}{2},\ \ \tau>0,
\end{equation*}
\begin{equation}
\left.
\begin{aligned}
v_{yy}-2\left(v-\frac12\right)^3&=L_m S^j\dot S,\\
v-\frac12&=P_m S^j
\end{aligned}
\right\}
\quad\text{on}\quad y=\gamma(\tau+a)^{1/3}+\frac{3\tau}{2},\ \tau>0,
\tag{3.5}
\end{equation}
and
\begin{equation*}
\left[v_{yy}-2\left(v-\frac12\right)^3\right]_{y=3\tau/2}=H_0(\tau+a)^k,
\qquad \tau>0,
\end{equation*}
$$ S(0)=S_0,$$
wherein $S(\tau)=\gamma(\tau+a)^{1/3}$.
The preceding system (3.5) was obtained in  \cite{ref31}  by setting
\begin{equation}
x=-\frac{3}{2}\tau+y,\qquad t=\tau,\qquad v=\frac12+u,
\tag{3.6}
\end{equation}
in the class of mKdV moving boundary problems
\begin{equation*}
u_t-6u^2u_x+u_{xxx}=0,\qquad 0<x<S(t),\ t>0,
\end{equation*}
\begin{equation}
\left.
\begin{aligned}
u_{xx}-2u^3&=L_m S^i,\dot S,\\
u&=P_m S^i,
\end{aligned}
\right\}
\quad\text{on}\quad x=S(t),\ t>0,
\tag{3.7}
\end{equation}
\begin{equation*}
\left[u_{xx}-2u^3\right]_{x=0}=H_0(t+a)^k,\qquad t>0,
\end{equation*}
\begin{equation*}
{S(0)=S_0 ~~ \textstyle{\rm with} ~~S(t)=\gamma (t+a)^{1/3}~.}
\end{equation*}

This nonlinear system has been shown to admit exact solution via Painlev\'e II
symmetry reduction \cite{ref12}.
In particular it admits the exact solution
\begin{equation*}
u= -\delta (t+a)^{-1/3}\phi'\left({{ x}\over{\epsilon(t+a)^{1/3}}}\right) 
\left[\phi\left({{x}\over{\epsilon(t+a)^{1/3}}}\right) \right]^{-1},
\end{equation*}
where $\phi$ is governed by an Airy  equation.
A novel class of exactly solvable moving boundary problems for the extended solitonic cuspon equation
results via the action of the reciprocal transformation (3.3) on the  Gardner moving 
boundary system (3.5).
The reciprocal extended cuspon moving 
boundary system  inherits the property of admittance of exact solution via Painlev\'e II
symmetry reduction.

\subsection*{Acknowledgements}

S.C.\ acknowledges  the Italian National Institute of Nuclear Physics (CSN4- MMNLP - Mathematical methods of nonlinear Physics- RM1), the Italian National Group of Mathematical Physics (GNFM--INdAM), SBAI Dept. of Sapirenza University of Rome, Italy and the project CTE of Regione Lazio,  for supporting the present research activity.

\label{lastpage}
\end{document}